\documentclass[12pt]{article}

\usepackage{natbib}
\usepackage{amssymb}
\usepackage{amsmath}
\usepackage{amscd}
\usepackage{latexsym}
\usepackage[dvipdfm,pdfstartview=FitH,colorlinks=true,citecolor=blue]{hyperref}
\newcommand{\hpc}{\hspace{0.1mm}}

\newcommand{\be}{\begin{equation}}
\newcommand{\ee}{\end{equation}}
\newcommand{\brr}{\begin{array}}
\newcommand{\ber}{\end{array}}
\newcommand{\bsf}{{\Psi({\bf x},{\bf q},t)}}
\newcommand{\bsfs}{{\left|\Psi({\bf x},{\bf q},t)\right|^2}}
\newcommand{\wfqt}{{\psi({\bf q},t)}}

\newcommand{\idwf}{\Psi(\bfx,\bfq_1,t;\bfx,\bfq_2,t)}

\newcommand{\bfx}{{\bf x}}

\newcommand{\bfq}{{\bf q}}
\newcommand{\dfq}{{\rm d}{\bf q}}
\newcommand{\dfx}{{\rm d}{\bf x}}
\newcommand{\psii}{{\psi_i}}

\newcommand{\ketpsis}{{\Psi_s}}
\newcommand{\ketpsia}{{\Psi_a}}
\newcommand{\permu}{{\hat{P}_{12}}}

\newcommand{\idv}{ideal individual}
\newcommand{\idvs}{ideal individuals}

\newcommand{\hvi}{\harvarditem}

\begin{document}

\begin{center}
{{\Large {\bf Identical ideal individual hypothesis}} \\ [5mm]
Jiao-Kai Chen } \\ [3mm]

{\it School of Physics and Information Engineering,\\ Shanxi Normal University, Linfen 041004, P. R. China \\ E-Mail: chenjk@sxnu.edu.cn, chenjkphy@outlook.com}
\end{center}

\vskip 1cm
\begin{abstract}
{The identical ideal individual hypothesis is proposed. According to this hypothesis, the identical ideal individuals should be classified into two classes: the bosonic individuals and the fermionic individuals. The bosonic individuals can occupy the same behavior state while the fermionic individuals can not be in the same behavior state.
We propose that human beings and many species of animals are fermionic, which can not occupy the same behavior state according to the Pauli exclusion principle.
An unified theoretical explanation is given for the natures of two important and seemingly irrelated phenomena in psychology: the existence of the personal space and the behavior differentiation under high population density condition.}
\end{abstract}

\vskip 5mm

\section{Introduction}
The application of concepts and principles of quantum theory in physics to psychology which we call quantum psychology here for simplicity is promising and is attracting more and more attention \citep{Aerts09qm,aerts94,Aerts13qm,atmanspacher02q,Bagarello19,bagarello18qf,bagarello18hb,bordely98hu,
Bruza15qm,Busemeyer12bksch,busemeyerqps,chen19psy,haven13sci,Khrennivov99sch,Melkikh19sch,Pothos09sch,
Rossi94qp,Triffet96sch,valle89qp}. Quantum psychology has been used to investigate many puzzling phenomena and problems accumulated in psychological science which can not be solved by classical theories \citep{Busemeyer12bksch}{\hpc}. Based on the ideal individual model, behavior coordinate system and quantum probability \citep{chen19psy}{\hpc}, we assume that the behavior state of an ideal individual should be represented by a behavior state function and then propose the identical ideal individual hypothesis.

The existence of the personal space \citep{Hall66ps,Katz37ps,Sommer59ps,Stern38ps,Uexhull57ps} and the behavior differentiation under high population density condition \citep{Calhoun62pd,Evans79pd,Marsden72pd}are two important and confusing issues in psychology. The natures of these two phenomena remain mysterious and irrelated in old theories.
We suggest that these two phenomena are quantum effects. By applying the identical {\idv} hypothesis, we offer a possible theoretical explanation in an unified approach for the natures of these two seemingly irrelated phenomena.

In the remainder of this paper we first introduce the concepts of the behavior coordinate system and the behavior state function. Second, we propose the identical ideal individual hypothesis. Third, we discuss the consequences of the identical ideal individual hypothesis. We offer a possible theoretical explanation in an unified approach for the natures of the existence of the personal space and the behavior differentiation under high population density condition.  Fourth, the conclusions are given.

\section{Preliminaries}\label{sec:behav}
\subsection{Behavior coordinate system}
As position of a point particle is the basic variable in Newtonian mechanics, behavior or behavior position of an individual is the basic variable in psychology. Behavior refers to the observable actions of an ideal individual, e.g. speech, body movement and emotional expression. (In this work, the case of human beings is discussed, other species can be investigated similarly.)
Except for the intrinsic quantities such as age and gender, quantities such as motivation, emotion and personality are supposed to be functions of behavior and the derivative of behavior with respect to time. In other words, these quantities can be revealed by behavior and its derivative \citep{tolman48}{\hpc}.

An ideal individual is an idealization of humans or other species of animals \citep{chen19psy}{\hpc}. For an ideal individual, every possible behavior can occur with equal probability. These possible behaviors are elements of the set whose members are the behaviors of humans and animals that have occurred in the past, are occurring at present and will occur in the future. Different behaviors show different properties and make distinction among different species and different individuals.

Let the behavior coordinate system be a coordinate system that specifies each behavior point uniquely in a behavior space by a set of numerical behavior coordinates \citep{hock15ft,rosenhan73ax}{\hpc}. The reference lines are $q_1$-axis, $q_2$-axis, $q_3$-axis and so on. To describe the behaviors of human beings, we can assume that
$q_1$-axis is the speech-axis, $q_2$-axis is the body-movement-axis and $q_3$-axis is the emotional-expression-axis.
For simplicity, we assume the behavior coordinate system is a Cartesian coordinate system, and the behavior coordinate space is a n-dimensional Euclidean space $\mathbb{R}^n$.
Although the behavior coordinate system is far from being well established now, it is useful for solving many problems and it can give some enlightening results.

\subsection{Behavior state function}
The behavior state of an ideal individual is assumed to be represented by a behavior state function $\bsf$ \citep{chen19psy}{\hpc} which represents one behavior pattern. $\bsf$ is a vector in the Hilbert space, where $t$ is time, $\bfx$ denotes the spatial space vector $\bfx=(x,y,z)$. The spatial position in psychology is some different from that in physics because an organism has volume and cannot be regarded as a particle in psychology although it can be regarded as a particle in physics which is defined to be an object of insignificant size. $\bfq$ denotes the behavior space vector $\bfq=(q_1,q_2,q_3,\cdots)$. $q_1$ is the speech coordinate, $q_2$ is the body-movement coordinate, $q_3$ is the emotional-expression coordinate, and $\cdots$ denotes other coordinates.

Adopting the Born hypothesis which gives the probability interpretation of the wave function in quantum mechanics, the behavior state function $\bsf$ is assumed to be a probability amplitude, such that $\bsfs$ gives the probability of finding an ideal individual at a behavior point $\bfq$ at time $t$ and spatial point $\bfx$. There is the normalization condition
\be
\int\bsfs \dfq\dfx=1.
\ee
The integration is over whole behavior space and spatial space.
In many cases, the spatial coordinates have small or even no effects on an individual's behavior and then the spatial coordinates can be neglected. In some cases, the spatial space is taken as part of the environment. In consequence, the behavior coordinates are highlighted in these cases and the spatial coordinates can be integrated out,
\be
\wfqt=c\int{\bsf}{\dfx}, \quad \int\psi^{\ast}(\bfq,t)\psi(\bfq,t)\dfq=1,
\ee
where $c$ is a normalization factor.
We assume the superposition principle holds for $\Psi$ and $\psi$,
\be
\Psi=\sum_{i=1}^{n}a_i\Psi_i,\quad \psi=\sum_{i=1}^{n}b_i\psii.
\ee
In many references \citep{Busemeyer12bksch,haven13sci}{\hpc}, the behavior state function $\wfqt$ has been used actually to discuss problems although the behavior variable $\bfq$ is not given explicitly.

\section{Identical ideal individuals}
Human beings and animals are macroscopic objects whose mechanical motion in spatial space is described not by quantum mechanics but by Newtonian mechanics. The considered quantum effects are not on mechanical motion but on behaviors. The ideal individuals can be regarded as being identical approximately under some conditions and then some quantum effects will become prominent. For simplicity, let $\bfx_1=\bfx_2=\cdots=\bfx$, $t_1=t_2=\cdots=t$, where $\bfx_i$ is the spatial position of the $i$-th individual at time $t_i$.

Two ideal individuals are identical if there should be no experiment that detects any intrinsic difference between them. The identical {\idvs} are those individuals that have the same age, gender, etc. and behave in the same manner under equal conditions.
Suppose there is a system of two indistinguishable individuals $1$ and $2$. Because of the identity of the ideal individuals, the states of the system obtained from each other by merely interchanging the two individuals must be completely equivalent. This means that, as a result of this interchange, the state function of the system can change only by an unimportant phase factor. This operation is carried out by the operator $\permu$
\be
\permu\idwf=\Psi(\bfx,\bfq_2,t;\bfx,\bfq_1,t)={\rm e}^{{\rm i}\lambda}\idwf,
\ee
where $\idwf$ is the behavior state function, $\lambda$ is some real constant. $\permu$ is the permutation operator,
\be
\permu=\permu^{-1}, \quad \permu\permu=I,\quad \permu^{\dag}=\permu,
\ee
where $I$ is the identity operator. A second exchange of these two individuals recreates the original state. Hence,
\be
\permu^2\idwf={\rm e}^{2{\rm i}\lambda}\idwf=\idwf,
\ee
yielding
\be
{\rm e}^{{\rm i}\lambda}=\pm1.
\ee
Therefore either
\be
\permu\ketpsis=\ketpsis\;\; {\rm or}\;\; \permu\ketpsia=-\ketpsia\;\; {\rm holds},
\ee
where
\be\label{asystate}
\ketpsis=\frac{1}{\sqrt{2}}(1+\permu)\idwf,\;\;
\ketpsia=\frac{1}{\sqrt{2}}(1-\permu)\idwf.
\ee
We call the behavior state function $\ketpsis$ with the eigenvalue $+1$ symmetric and $\ketpsia$ with the eigenvalue $-1$ antisymmetric with respect to the exchange of two indistinguishable ideal individuals.

Whether individuals are described by a symmetric or an antisymmetric behavior state function will depend on their nature. The identical ideal individuals should be classified into two classes: the bosonic individuals and the fermionic individuals. The bosonic individuals can be described by a symmetric behavior state function $\ketpsis$ and they can occupy the same behavior state. The fermionic individuals are described by an antisymmetric behavior state function $\ketpsia$ and they can not be in the same behavior state. If the fermionic individuals occupy the same behavior state, we can obtain $\ketpsia=0$ from Eq. (\ref{asystate}), which we call ``zero state" or ``dead state". This result is known as the Pauli exclusion principle which is sometimes expressed by stating that no two fermionic individuals can occupy the same behavior state.

\section{Discussions}
If the difference between the spatial positions and some intrinsic properties of individuals are indistinguishable approximately or can be neglected to some extent as the ideal individuals are congregated with high population density \footnote{The population density is the number of organims per unit volume, $n=N/V$, where $N$ is the number of organisms, $V$ is the spatial volume occupied by organisms. The population density is high if $n{\ge}n_{c}$, where $n_{c}$ is the critical value of population dentisy. $n_{c}=1/V_{ps}$, where $V_{ps}$ is the personal space of this species of organisms. Under high population density condition, the organisms can be regarded as being indistinguishable by their spatial positions in quantum psychology.} , these ideal individuals are assumed to be identical.

If the identical ideal individuals are bosonic, they can occupy the same behavior state and a large fraction of them can occupy the lowest behavior state under some conditions as a Bose-Einstein condensate. It can be expected that the bosonic individuals will congregate not only in behavior space but also in spatial space under some conditions. Evidently, the bosonic individuals who have the same behavior state and are assembled with high population density are not in the advantageous position in competition and survival of species according to Darwin's theory of nature selection because they as a species are vulnerable in suddenly appeared harsh environments. We conjecture that the species whose members are bosonic will be small in number if there exist this kind of species in nature and their living environment will be very different from the environment where the fermionic species live.

If the identical ideal individuals are fermionic, they can not occupy the same behavior state according to the Pauli exclusion principle. The fermionic individuals taking the same behavior state or behavior pattern are repulsive. Consequently, we can conjecture that the fermionic individuals need their personal spatial space to be distinguished by spatial positions and then they are not identical again so that they can behave not only in different patterns but also in the same pattern. This discussion offers a possible theoretical explanation for the existence of the personal space \citep{Sommer59ps}{\hpc} which is a quantum effect in spatial space. Simultaneously, human beings and many species are expected to be fermonic. According to the above discussions, we also suggest that not only human beings but also other species of organisms have their personal spaces if their behavior states can be described by the behavior state function.

The fermionic individuals occupying the same behavior state will be in ``zero state", otherwise, they will be forced by the Pauli exclusion principle to be in different behavior states or to form different behavior patterns. This discussion gives a possible theoretical explanation for the behavior differentiation under the high population density condition \citep{Calhoun62pd}{\hpc}. The behavior differentiation is a quantum effect in behavior space.
Let the behavior state be represented by the behavior state function $\Psi_{nm}$, where $n$ and $m$ are quantum numbers. For degenerate states, there are many possible values of $m$ for a given value of $n$. Therefore, some rats can behave in the same behavior pattern with different $m$ which does not violate the Pauli exclusion principle. It offers a possible explanation for the quantities of rats exhibiting the same behavior pattern under high population density condition \citep{Calhoun62pd}{\hpc}.

The existence of the personal space and the behavior differentiation under high population density condition seem irrelated and the natures of them has been remaining mysterious ever before. Now we tend toward believing that they are quantum phenomena arising from the identity of the ideal individuals.
The unified explanations for the natures of these two phenomena can be regarded as an evidence that psychology can be described by  quantum theory. More prominent quantum effects are expected to be observed in the behaviors of simple organisms if their behaviors can be described by quantum theory. Experiments on simple organisms such as amoeba, paramecium and bacteria under high population density condition are expected.

\section{Conclusions}
Based on the behavior coordinate system, the ideal individual model and quantum probability, we assume that the behavior state of an ideal individual should be represented by a behavior state function. Then we propose the identical ideal individual hypothesis. According to this hypothesis, the identical ideal individuals should be classified into two classes: the bosonic individuals and the fermionic individuals. The bosonic individuals can occupy the same behavior state while the fermionic individuals can not be in the same behavior state. We conjecture that the bosonic species will be small in number or be very different from the fermionic species if they exist in nature.
We propose that human beings and many species of animals are fermionic, which can not occupy the same behavior state according to the Pauli exclusion principle.

The existence of the personal space and the behavior differentiation under high population density condition are explained theoretically in an unified approach by using the identical {\idvs} hypothesis. The existence of the personal space is a quantum effect in spatial space caused by the identity of the {\idvs} while the behavior differentiation under high population density condition is a quantum effect in behavior space.


\end{document}